\documentclass[prl,reprint,showpacs,twocolumn,superscriptaddress,floatfix]{revtex4-1}
\usepackage{epsfig}
\usepackage{overpic}
\usepackage{epsfig}

\usepackage{amsmath}
\usepackage{amssymb}
\usepackage{amsfonts}
\usepackage{mathptmx}
\usepackage{dcolumn}
\usepackage{eucal}
\usepackage{bm}
\usepackage{color}
\usepackage[colorlinks,linkcolor=blue,citecolor=blue]{hyperref}
\usepackage{epstopdf}
\usepackage{bbold}
\usepackage{url}

\usepackage{dsfont}

\def\>{\right\rangle}
\def\<{\left\langle}
\def\be{\begin{equation}}
\def\ee{\end{equation}}
\def\ba{\begin{array}{lll}}
\def\ea{\end{array}}

\def\beq{\begin{eqnarray}}
\def\eeq{\end{eqnarray}}

\begin{document}
\title{Optimal work-to-work conversion of a nonlinear quantum brownian duet}
\author{ Matteo Carrega}
\email{matteo.carrega@nano.cnr.it}
\affiliation{NEST, Istituto Nanoscienze-CNR  and Scuola Normale Superiore, I-56127 Pisa, Italy}
\author{ Maura Sassetti}
\affiliation{Dipartimento di Fisica, Universit\`a di Genova, Via Dodecaneso 33, 16146 Genova, Italy}
\affiliation{SPIN-CNR, Via Dodecaneso 33, 16146 Genova, Italy}
\author{ Ulrich Weiss}
\affiliation{II. Institut f\"ur Theoretische Physik, Universit\"at Stuttgart, D-70550 Stuttgart, Germany}
\begin{abstract}
Performances of work-to-work conversion are studied for a dissipative nonlinear quantum system with two isochromatic phase-shifted drives.
It is shown that for weak Ohmic damping simultaneous maximization of efficiency with finite power yield and low power fluctuations can be achieved. Optimal performances of these three quantities are accompanied by a
shortfall of the trade-off bound recently introduced for classical thermal machines. This bound can be undercut down to zero for sufficiently low temperature and weak dissipation,
where the non-Markovian quantum nature dominates.
Analytic results are given for linear thermodynamics. These general features can persist in the nonlinear driving regime near to a maximum of the power yield and a  minimum of the power fluctuations. This broadens the scope
to a new operation field beyond linear response.
\end{abstract}
\maketitle
{\it Introduction. ---}
Major efforts in  classical and quantum thermodynamics are directed at strategies to efficiently manipulate and transform varied forms of energy into useful ones~\cite{seifert12, kosloff13, benenti17, esposito09, campisi11, levy12, carrega15, carrega16, ludovico14, taddei18, arrachea19}. Optimal heat to work conversion is a founding principle for a wide range of applications, from industrial processes to biological functionalities, thermoelectricity and photovoltaics~\cite{benenti17}.
The seminal work of Carnot established an upper bound, $\eta\leq \eta_{\rm C}$, for the efficiency $\eta$
of all heat engines. It is argued that this bound is saturated for reversible operation with  vanishing power yield~\cite{benenti17, curzon75, whitney1, whitney2, broeck05, goychuk13}.
This poses a severe restriction, as finite power output is essential for useable  thermal machines.
However, both efficiency and yield have to be sufficiently large for a well working engine: if $\eta$ is small, a major part of energy is wasted, while low output power would not supply sizeable work in finite time.
Various studies  focussed on the  maximum attainable efficiency at a given finite power
yield~\cite{holubec15, ryabov16, ma18, esposito10, cavina17}.
General relations linking maximum power, maximum efficiency and minimum dissipation have been derived  within linear thermodynamics ~\cite{benenti11, proesmans16, proesmans16duet, proesmans17, benenti18}.
It has been proposed that constraints on efficiency at finite power could be overcome in specific settings, e.g.,
by    breaking time-reversal symmetry \cite{benenti11}.
Various attempts to get close to Carnot efficiency upon retaining finite power have been
made~\cite{allahverdyan13, campisi16, holubec17, polettini17}, e.g., by suggesting working
points near to critical phase transitions~ \cite{campisi16, holubec17}.
However, these settings are  impaired by large power fluctuations, which  undermine effective working
of the machine~\cite{holubec17, solon18}.

A universal trade-off  criterion,  a bound  constraining these quantities and holding for a wide class
of classical Markovian  systems operating  in the steady state, has been
derived~\cite{ barato15,shiraishi16, horowitz17, pietzonka18, holubec18, shiraishi18}.
The bound implies that high power yield, efficiency close to the Carnot value, and small power fluctuations are not compatible.
Generalizations of the trade-off bound holding for time-periodic systems have been
discussed~\cite{barato18, koyuk18}.
Recently, weakening of the trade-off bound  has been found in ballistic multi-terminal transport
\cite{brandner18} and in coherent electron transport through resonant single- and double-dot junctions without~\cite{agarwalla18} and with electron interactions~\cite{coherence18}.

It is therefore interesting to  study systematically the impact of quantum effects on the trade-off criterion
in a key model of quantum transport:  a quantum Brownian particle (QBP) moving in a tight-binding (TB) lattice and coupled to a thermal reservoir creating Ohmic friction.
With two time-dependent external drives, the system forms a quantum Brownian duet and acts as
an iso-thermal work-to work converter.  The QBP  model has widely varied applications~\cite{weissbook}. It  describes, e.g. the current-voltage characteristics of a Josephson junction~\cite{schoen90, andersen13, wendin17}, transport of charge through impurities in quantum wires~\cite{kane92, voit}, and tunneling of edge currents through constrictions in one-dimensional interacting fermion systems~\cite{wen90, kane95, giamarchi2004, ferraro14, dolcetto16}. The Ohmic spectral coupling entails power laws for the temperature and bias dependence of tunneling rates. This leads for weak damping, e.g., to increasing tunneling with decreasing temperature~\cite{grabert85,fisher85b}.

In this Letter we first show within linear thermodynamics and weak tunneling, that the specific Ohmic features make possible to optimize performance upon simultaneous adjustment of large power yield, high efficiency and low power fluctuations.
We find that for weak damping the trade-off quantity can fall below the classical bound and even can
approach zero, as temperature is decreased and the non-Markovian quantum regime is reached.
We also focus on a hitherto mostly disregarded regime beyond linear thermodynamics, in which
nonlinear external driving and response prevails. We there discover a parameter regime with sizeable power yield,
low power fluctuations and efficiency still close to unity.

{\it Model.---}
Consider a quantum brownian particle (QBP) in a tight-binding (TB) lattice bilinearly coupled to  a thermal bath of harmonic oscillators at inverse temperature $\beta$. The TB-Hamiltonian is
$H_{\rm TB} = -\frac{1}{2} \hbar\Delta\sum_n(a_n^\dagger a_{n+1}^{}+{\rm h.c.})$ and
the bath-plus-coupling term is
$ H_{\rm RI} = \sum_\alpha[\frac{p_\alpha^2}{2m_\alpha}+ \frac{m_\alpha\omega_\alpha^2}{2}(x_\alpha-\frac{c_\alpha}{m_\alpha\omega_\alpha}q)^2] $, where $q=q_0\sum_na_n^\dagger a_n^{}$, and $q_0$ is the lattice constant.
The spectral bath coupling is
$J(\omega) = \frac{\pi}{2}\sum_\alpha
\frac{c_\alpha^2}{m_\alpha\omega_\alpha}\delta(\omega-\omega_\alpha)$~\cite{leggett87,weissbook}.
In the Ohmic scaling limit we have  $J(\omega)=2\pi K \omega$, where $K$ is the  dimensionless damping strength. The bare transfer amplitude $\Delta$ is adiabatically renormalized by the modes $\omega>\omega_{\rm c}$ to the dressed amplitude $\Delta_{\rm r} = \Delta (\Delta/\omega_{\rm c})^{K/(1-K)}$.
{ The QBP model maps inter alia on quasiparticles tunneling through a quantum point contact (QPC) in the fractional quantum Hall (FQH) regime~\cite{wen90, kane95}, whereby $K$ corresponds to the  fractional filling factor $\nu$.} The 
weak-damping regime $K \ll 1$ matches up with  strong repulsive short-range electron interactions.

Here we study energy in- and output of the QBP under time-periodic drive
$H_{\rm ext}(t) = -\hbar\epsilon(t) \,q/q_0$, where $\epsilon(t) = \epsilon(t+{\cal T})$.
{When the QBP is subjected to two independent drives, $\epsilon(t)=\epsilon_1(t)+\epsilon_2(t)$, 
as discussed for a classical setting in Ref.~\cite{proesmans16duet, proesmans17},
it can operate as a work-to-work converter. But it will require that the respective work rates or powers,
$\dot{W}_i\equiv P_i(t)=\epsilon_i(t)\dot{q}(t)$, $i=1,\, 2$, can be distinguished.}
We now choose $\hbar=k_{{\rm B}}=q_0=1$.

At long times, the power $P_i(t)$ reaches the periodic state $\overline{P}_i(t) = \overline{P}_i(t+{\cal T})$,
and the mean power is $\langle P_i\rangle = \int_0^{\cal T} dt \, \overline{P_i}(t)/{\cal T}$.
 With the deviation $\delta P_i(t) = P_i(t) -{\overline P}_i(t)$, the power fluctuations are
$\overline{D}_i(t)=\int_0^t dt'\int_0^t dt''\,\overline{\delta P_i(t')\delta P_i(t'')}/t$, and the mean power spread is
$\langle D_i\rangle = \int_0^{\cal T} dt\, \overline{D}_i(t)/{\cal T}$.   

Consider now mean power and power fluctuations of the driven QBP. First, we  deal with order $\Delta^2$,
which is the leading contribution in the weak-tunneling regime. It
describes transport via nearest-neighbor transitions,  and single-electron transport in the related fermionic model.
We have~\cite{sm}
\begin{align}\label{eq:meanpower}
{\langle P_i\rangle} &=
 \int_0^{\infty} \!\!d\tau~k_{\rm P}(\tau)\int_0^{\cal T}\!\!
\frac{dt}{\cal T}\,\epsilon_i(t)\sin[G(t,t-\tau)]\, ,    \\  \label{eq:vardef}
\langle { D}_{i}\rangle &= \!\!\int_0^{\infty}\!\!d\tau\,k_D(\tau)  \!\!\int_0^{\cal T} \!\!\frac{dt}{\cal T}~\epsilon_i(t)\epsilon_i(t-\tau)\cos[G(t,t-\tau)]\, ,
\end{align}
with $G(t_2,t_1)=\int_{t_1}^{t_2}dt'\, [\epsilon_1(t')+\epsilon_2(t')]$. The functions $k_{\rm P/D}(\tau)$
carry the  amplitude factor  $\Delta_{\rm r}^{2-2K}$ and the Ohmic bath correlations. They read
$k_{\rm D}(\tau)=\cot(\pi K) k_{\rm P}(\tau)$, and~\cite{leggett87,weissbook, sassetti92}
\be
\label{eq:ks}
k_{\rm P}(\tau)= \Delta_{\rm r}^2 \Big(\frac{\pi}{\beta\Delta_{\rm r}}\Big)^{2 K}\frac{\sin(\pi K)}{\sinh(\pi\tau/\beta)^{2 K}} \, .
\ee
When the  mean powers have opposite sign, $\langle P_1\rangle \langle P_2\rangle <0$, the QBP entity is
acting as  work-to-work converter with the positive power being the {input, and the negative power being the output or  yield~\cite{seifert12, proesmans17}}.
If  $\langle P_2\rangle$ is the {input} and $\langle P_1\rangle$ is the yield, the efficiency of the converter is $\eta \equiv |\langle P_1\rangle |/\langle P_2\rangle\leq 1$.
Optimal performance is characterized by maximal efficiency at given input. However,  optimization of the converter should also conform to power fluctuations as low as possible.
The latter may be rated with the estimate of  relative uncertainty
\be
\label{eq:sigmapower}
\Sigma_1=\sqrt{\langle D_{1}\rangle/\langle P_1\rangle^2} \, .
\ee
It has been argued and proven for a huge class of steady-state heat engines with internal classical states that
there is a trade-off between large power, high efficiency and low relative uncertainty, being expressed by the joint
bound~\cite{barato15,shiraishi16, horowitz17, pietzonka18, shiraishi18}
\be \label{eq:tradeoff}
Q_1\equiv \beta   |\langle P_1\rangle| (1/\eta-1 )\Sigma_1^2  \ge 2 \, .
\ee
If efficiency is close to unity with  considerable yield, the bound implies
that the power fluctuations are quite large. Conversely, if the bound is broken, simultaneous attainment of maximal efficiency, sizeable yield and low power fluctuations are within reach. This can happen indeed, as shown below.

{\it Work-to-work conversion with two monochromatic drives.---}
{If the frequencies of the two drives would be different,
the work rates could clearly be distinguished. But here we choose the same frequency, since otherwise  the converter could not operate in the linear regime, as different frequencies would not couple~\cite{proesmans16}.} We put
\beq
\epsilon_1(t)&=& F_1\sin(\omega t)  \, ,     \nonumber \\   \label{eq:drive1}
\epsilon_2(t)&=& F_2\cos(\omega t-\varphi)   \, .
\eeq
Here, the tunable phase shift $\varphi=\arctan\alpha$ determines the (time-reversal) asymmetry between the two drives.

{The drives (\ref{eq:drive1}) can be combined into
\beq \nonumber
\epsilon(t) &=& {\cal F} \sin(\omega\,t+\Phi)\, , \\   \label{eq:drive2}
{\cal F} &=& {\rm sgn}\,(F_1+F_2 \sin\varphi) [F_1^2+F_2^2+2F_1 F_2\sin\varphi]^{1/2}\, , 
 \\ \nonumber 
\Phi &=& \arctan[F_2 \cos\varphi/(F_1+F_2 \sin\varphi)]  \, .
\eeq               }
With (\ref{eq:drive1}) and  (\ref{eq:drive2})  the time-averaged powers are found as~\cite{sm}
\beq   \label{eq:pfull}
\langle P_1\rangle &=& \int_0^{\infty} \!\!\!d\tau\;k_{\rm P}(\tau)\,F_1\, J_1[A(\tau)]\cos(\Phi-\omega\tau/2)\, ,   
\\  \nonumber
\langle P_2\rangle &=& \int_0^{\infty} \!\!\!d\tau\;k_{\rm P}(\tau)\,F_2\, J_1[A(\tau)]\sin(\Phi+\varphi-\omega\tau/2)
\eeq
with $A(\tau) = 2 {\cal F}\sin(\omega\tau/2)/\omega$, and $J_n(z)$ is a Bessel function.
The mean power fluctuations are  found from Eq.~(\ref{eq:vardef}) as
\beq
\langle D_{i}\rangle &=& \frac{1}{2} \int_0^{\infty}d\tau\;k_{\rm D}(\tau)\, d_i^{}(\tau)\, ,
\qquad\quad i=1,2\nonumber \\   \label{eq:dfull}
d_1(\tau)&=& F_1^2\left\{ J_0[A(\tau)]\cos(\omega\tau) - J_2[A(\tau)]\cos(2\Phi) \right\} \, ,    \\    \nonumber
d_2(\tau)&=& F_2^2\left\{ J_0[A(\tau)]\cos(\omega\tau) + J_2[A(\tau)]\cos(2\Phi+2\varphi) \right\} \, .
\eeq
The expressions (\ref{eq:pfull}) - (\ref{eq:dfull}) are exact in the weak-tunneling limit for arbitrary strength of the driving amplitudes $F_1$ and $F_2$.

{As the drives (\ref{eq:drive1}) have common frequency, one may question
 whether the  powers (\ref{eq:pfull}) and fluctuations (\ref{eq:dfull}) can be experimentally distinguished.
This is possible, in fact, when the forces (\ref{eq:drive1}) are independent, e.g., when they operate spatially separated. A system, which can be mapped on 
the driven QBP, is a quantum point contact (QPC) in a fractional quantum Hall (FQH) bar {\cite{wen90, kane95, dubois2013, glattli, vannucciprl, ronettiprb}} with two spatially separated terminals at which the gate voltage drives are applied. The filling factor $\nu$ corresponds to the Ohmic coupling parameter $K$. In such physical implementation, 
the  powers (\ref{eq:pfull}), and power fluctuations (\ref{eq:dfull}) can be measured individually.
The QPC model with mapping on the QBP is discussed in the Supplemental Material~\cite{sm}. }

{\it Linear response. ---}
\begin{figure}
\centering
\begin{overpic}[width=0.495\columnwidth]{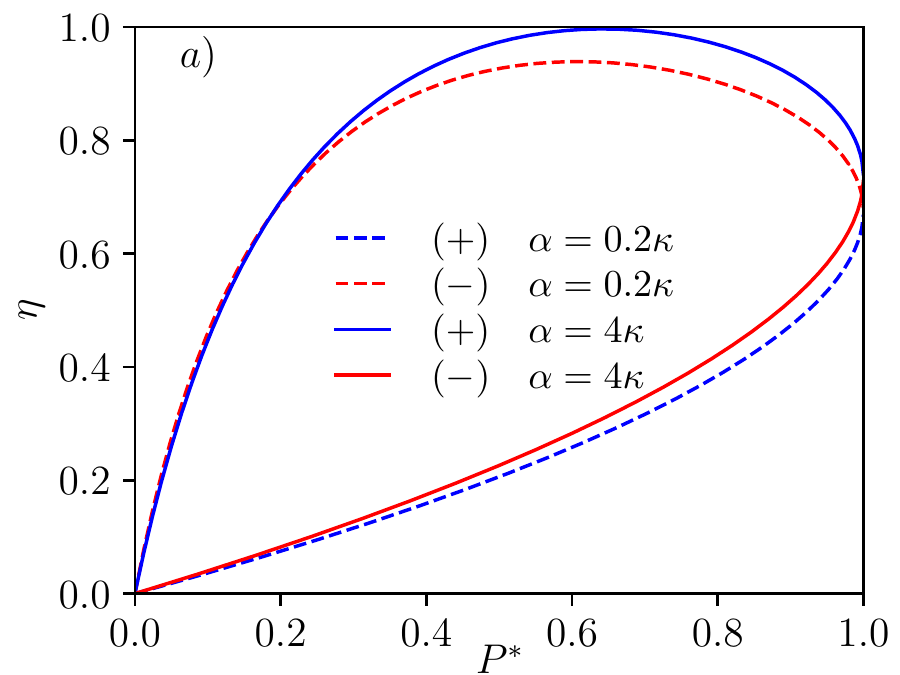}
\end{overpic}
\begin{overpic}[width=0.495\columnwidth]{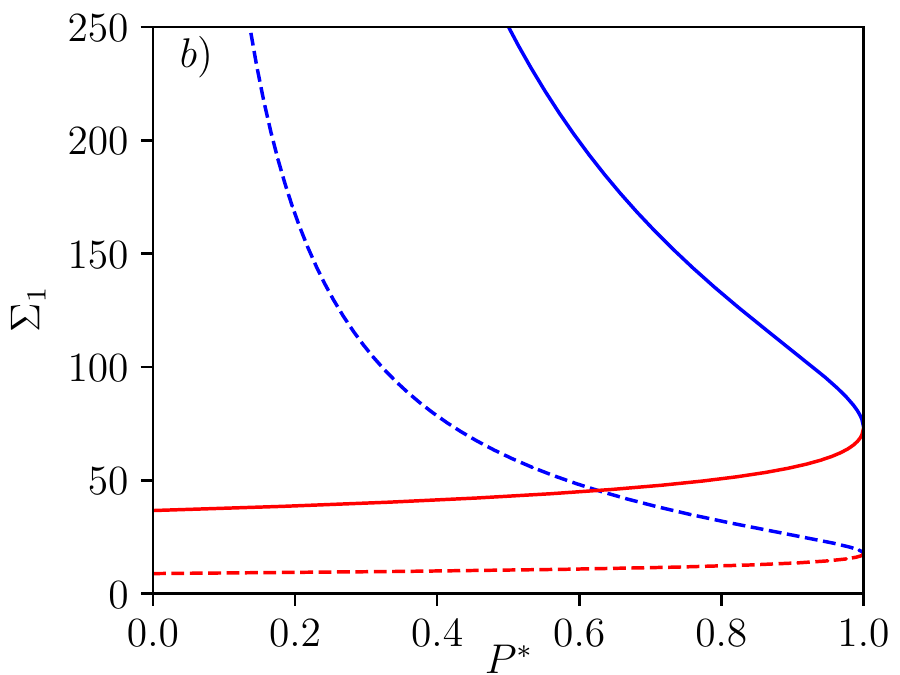}
\end{overpic}
\caption{
Efficiency (Panel a) and power fluctuations (Panel b) in the LR regime as function of $P^*$ (see text). Blue (red) color refers to the + (-) branch. Solid curve is $\alpha = 4 \kappa$ and dashed curve $\alpha= 0.2 \kappa$. Other parameters are $\beta\omega=6$, $\omega= 5\, \Delta_{\rm r}$, $\Delta_{\rm r}=\,1$, $K=0.1$, and $F_2=0.1 \omega $.  }
\label{fig:one}
\end{figure}
In the linear response (LR) regime, the dependence of the mean powers on the driving amplitudes $F_1$ and $F_2$ is expressed in terms of the Onsager matrix
${\cal L}$ as $\langle P_i\rangle = F_i \sum_{j=1,2}  {\cal L}_{i,j} \,F_j $. We get from Eqs.~(\ref{eq:pfull})
\beq \label{eq:onsmatrix}
{\cal L}= L_{\,\rm s}(\omega) \left(\begin{array}{cc}
1 & \frac{\alpha + \kappa}{\sqrt{1+\alpha^2}} \\
\frac{\alpha -\kappa}{\sqrt{1+\alpha^2}} & 1
\end{array}\right) \, ,
\eeq
where $\kappa= L_{\,\rm c}(\omega)/L_{\,\rm s}(\omega)$, and where
\beq
 L_{\,\rm s}(\omega) &=& \frac{1}{2\omega}\int_0^{\infty}d\tau~ k_{{\rm P}}(\tau)\sin(\omega \tau) \, ,  \nonumber \\
 L_{\,\rm c}(\omega) &=&  \frac{1}{2\omega}\int_0^{\infty}d\tau~k_{{\rm P}}(\tau)\,[1-\cos(\omega\tau )]  \, .
\eeq

The functions $L_{\,\rm s}(\omega)$ and $\kappa$ bear Ohmic
bath correlations. Dependence on $\beta$, $\omega$ and $K$ can be given in analytic form~\cite{sm}.
The  parameter $\alpha$ controls the phase shift of the drive (\ref{eq:drive1}). In addition, in the LR regime the power variance becomes $\langle D_i\rangle = F_i^2\omega\coth(\beta\omega/2) L_{{\rm s}}(\omega)$~\cite{sm}.

The Onsager matrix conveys the interplay of phase tuning of the driving forces and
exchange of energy $\omega$ between bath and QBP.
In the limit $\alpha \to \infty$ ($\varphi\to\pi/2$), the Onsager matrix is symmetric, and  the work-to-work converter operates time-reversal symmetrically.
As $\alpha$ is lowered, the Onsager matrix gets anti-symmetric admixtures, and  time-reversal symmetry is broken. In our setting, this scenario sets in without switching on external magnetic fields~\cite{benenti11}. In the limit $\alpha\to 0$ ($\varphi\to 0$), the Onsager matrix is antisymmetric. Upon tuning $\alpha$, one moves forth or back between these limiting cases.

For the linear model (\ref{eq:onsmatrix}), the maximum output power is at 
$F_1 = F_{1,\rm MP}\equiv -(\alpha+\kappa) F_2/(2\sqrt{1+\alpha^2})$,
\be\label{eq:pmax}
\langle P_{1,\rm MP}\rangle\equiv \langle P_1(F_{1,\rm MP},F_2)\rangle
 = - L_{\,\rm s}(\omega) \,  F_{1,\rm MP}^2   \, .
\ee
The condition $\langle P_1(F_1,F_2)\rangle /\langle P_{1,\rm MP}\rangle = P^\star$ has two roots, which are
$F_{1,\pm} =  \big(1\pm \sqrt{1- P^\star}\big) \,  F_{1,\rm MP}$ .
Correspondingly, efficiency and  power fluctuations as functions of $P^\star$ have two branches,
\beq
\label{eq:etabranches}
\eta_{\pm}^{} &=&\frac{P^*}{2}\frac{X}{1+2/Y\mp\sqrt{1-P^*}} \, , \\
\label{eq:sigmabranches2}
\Sigma_{1,\pm}&=& \frac{2}{F_2}
 \frac{\sqrt{1+\alpha^2}}{\alpha+\kappa}\sqrt{\frac{\omega\coth(\beta\omega/2)}{L_{\,\rm s}(\omega)} }
\frac{1\pm\sqrt{1-P^*}}{P^* }  \, ,
\eeq
where $X=[\alpha + \kappa)/(\alpha-\kappa)$ and $Y=(\alpha^2-\kappa^2)/ (1+\kappa^2)$.
The respective two branches collide at $P^\star=1$.

The two branches are plotted versus $P^\star$ in Fig.~\ref{fig:one} for  efficiency (left) and  power
fluctuations (right).
The behaviors are qualitatively different for $\alpha >\kappa$ and $\alpha <\kappa$.
The left panel shows that high efficiency can be reached on the (+)-branch when $\alpha>\kappa$, and
on the (-)-branch when $\alpha <\kappa$. In contrast, low power fluctuations arise only in branch (-) when
$\alpha < \kappa$. {Hence} high efficiency is compatible with low power fluctuations  when
$\alpha <\kappa$, i.e., when the antisymmetric off-diagonal parts of the Onsager matrix outweigh the symmetric ones.

To find out  optimum working conditions, we now focus on the maximum efficiency (ME) at notable power
yield. The efficiency $\eta(F_1)$ for fixed $F_2$ is maximal at $F_1= F_{1,\rm ME}$, where
$F_{1,\rm ME}= - F_2 (\sqrt{1+\kappa^2}-\sqrt{1+\alpha^2})/(\kappa-\alpha)$, and is given by
\be
\label{eq:etamelin}
\eta_{\rm ME}^{}= \frac{\sqrt{1+\alpha^2}-\sqrt{1+\kappa^2}}{\sqrt{1+\alpha^2}+\sqrt{1+\kappa^2}}\;\frac{\alpha
+\kappa}{\alpha-\kappa}~\, .
\ee

\begin{figure}
\centering
\begin{overpic}[width=0.495\columnwidth]{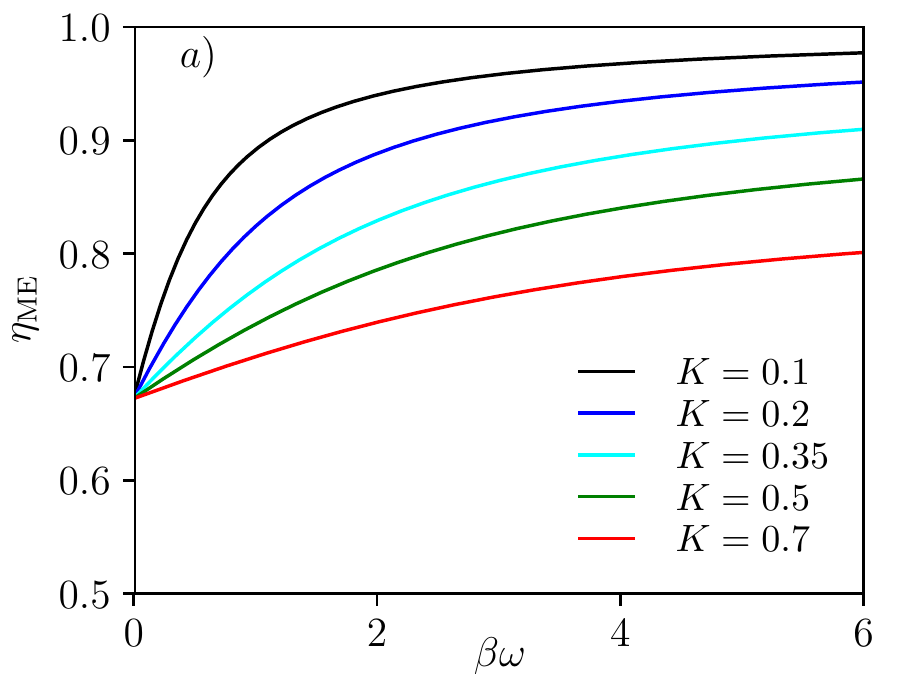}
\end{overpic}
\begin{overpic}[width=0.495\columnwidth]{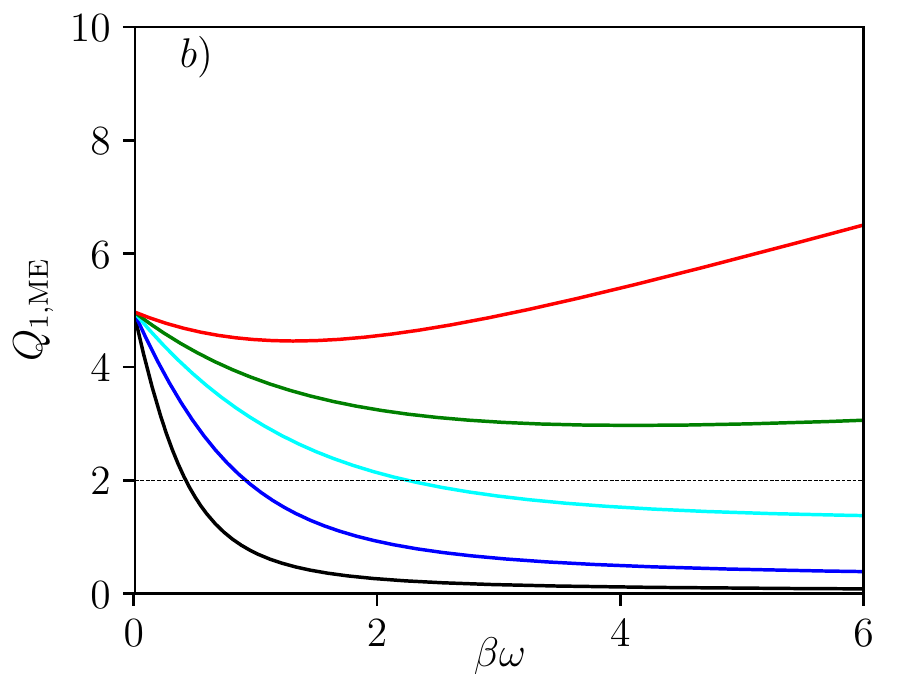}
\end{overpic}
\caption{
Maximum efficiency $\eta^{}_{\rm ME}$ (Panel a)  and trade-off criterion $Q_{1,\rm ME}$ (Panel b) versus $\beta\omega$ in the LR regime for $\alpha=5$ and different $K$ (both in a and b). See text.
Shortfall of the bound 2 occurs for $K \lessapprox 0.4$.
The gradual increase of $Q_{1,\rm ME}$ in the range $1/4<K\lessapprox 0.4$ takes place at higher $\beta\omega$
than shown in Panel b).  }
\label{fig:two}
\end{figure}

Fig.~\ref{fig:two}(a) shows $\eta_{\rm ME}$ versus $\beta\omega$ for different interaction strength $K$. As $K$ is decreased,
 $\eta_{\rm ME}$ is strictly increasing.
In the asymptotic non-Markovian low temperature regime $\beta\omega\gg 1 $, in which \cite{remark}
\be  \label{eq:kappaas}
\kappa(\beta\omega) \to \tan(\pi K)
\left[ \, (\beta\omega/(2\pi))^{1-2K}  \Gamma(K)^2/\pi -1 \,\right]   \, ,
\ee
the function $\kappa(\beta \omega)$ diverges as $(\beta\omega)^{1-2K}$ for $0<K<1/2$ and is a positive constant  for $1/2 <K<1$. Hence, as  $\beta\omega\to\infty$, $\eta_{\rm ME}^{}$ reaches unity in the former, and a value less than unity in the latter case. For $K\ll 1/2$, the prefactor of the term $(\beta\omega/(2\pi))^{1-2K}$ in Eq.~(\ref{eq:kappaas})
 is $\pi/K$. Thus, for weak Ohmic damping, or large repulsive Coulomb interaction in the associated fermionic transport model, the ME efficiency dwells close to unity in a considerably wide temperature range.

{
Eventually, with the function $c(x)= 2x \coth(x/2) $, the trade-off criterion (\ref{eq:tradeoff}) at $F_1=F_{1,\rm ME}$ takes the concise form~\cite{sm}
\be\label{eq:tradeoff1}
Q_{1,\rm ME} =c(\beta\omega)\frac{\sqrt{1+\alpha^2}}{(\alpha+\kappa)^2 }  \left(
\frac{1-\alpha\kappa}{\sqrt{1+\kappa^2}}+\sqrt{1+\alpha^2}\right)  \, .
\ee
}

In Fig.~\ref{fig:two}(b) the quantity $Q_{1,\rm ME}$ is plotted versus $\beta\omega$  for different values of $K$.
Since $\kappa(\beta\omega\to 0, K)\to 0$, the curves start out for all K at the value
$4\sqrt{1+\alpha^2}(1+\sqrt{1+\alpha^2})/\alpha^2$. In the regime $1/2<K<1$, we have
$\kappa(\beta\omega\to \infty)=-\tan(\pi K)$, and hence  $Q_{1,\rm ME}$ grows linearly with
inverse temperature at low temperatures, whereas $|\langle P_{1,\rm ME}\rangle|$ and $\Sigma_{1,\rm ME}$ become constant in this limit.
In contrast, in the range $0<K<1/2$, $\kappa(\beta\omega\to \infty)$ diverges asymptotically as $(\beta\omega)^{1-2K}$.  Thus,  the power  $|\langle P_{1,\rm ME}\rangle|$  grows
as $(\beta\omega)^{1-2K}$, the relative uncertainty  $\Sigma_{1,\rm ME}$ drops to zero as $(\beta\omega)^{2K-1}$,
and the quantity $Q_{1,\rm ME}$ varies as  $(\beta\omega)^{4K-1}$ in this limit.
As a result,  $Q_{1,\rm ME}$ diverges in the range $1/4<K<1/2$, stays flat below $2$ for $K=1/4$, and drops to zero when $K$ is in the range $0<K<1/4$, as $\beta\omega\to \infty$.
Hence the QBP work converter has optimal performance for weak damping $0<K<1/4$.
With decreasing temperature the  quantity $ Q_{1,\rm ME} $ falls well
 below the classical bound $2$, and  eventually drops to zero, as the non-Markovian quantum regime
is reached. Hence large power, high efficiency and small power fluctuations are in fact compatible.

{We have investigated the impact of an additional $n$-fold frequency drive in the output, 
$\epsilon_1(t) =F_1[\sin(\omega t+ \gamma_n \sin(n \omega t)]$. We found that the behaviors shown in
Figs.~ \ref{fig:one} and \ref{fig:two} change only marginally for $0<\gamma_n < 1$. Details are given in Ref.~\cite{sm}.      }

{\it Nonlinear response. ---}
The above results of the LR regime hold when the driving amplitude $F_2$ is sufficiently small,
$F_2 \ll \alpha\omega$.
For larger $F_2$, the interplay of nonlinear driving with bath correlations becomes significant, and the ME analysis must start with the original expressions (\ref{eq:pfull}) and  (\ref{eq:dfull}).  The ME point $F_1=F_{1,\rm ME}$ is found as numerical root of  $d\eta/dF_1=0$. With this, numerical  nonlinear response (NLR)  computation of
$P_{1,\rm ME}$, $\eta_{1,\rm ME}^{}$, $\Sigma_{1,\rm ME}$ and $Q_{1,\rm ME}$ is straight.
\begin{figure}
\centering
\includegraphics[width=\columnwidth]{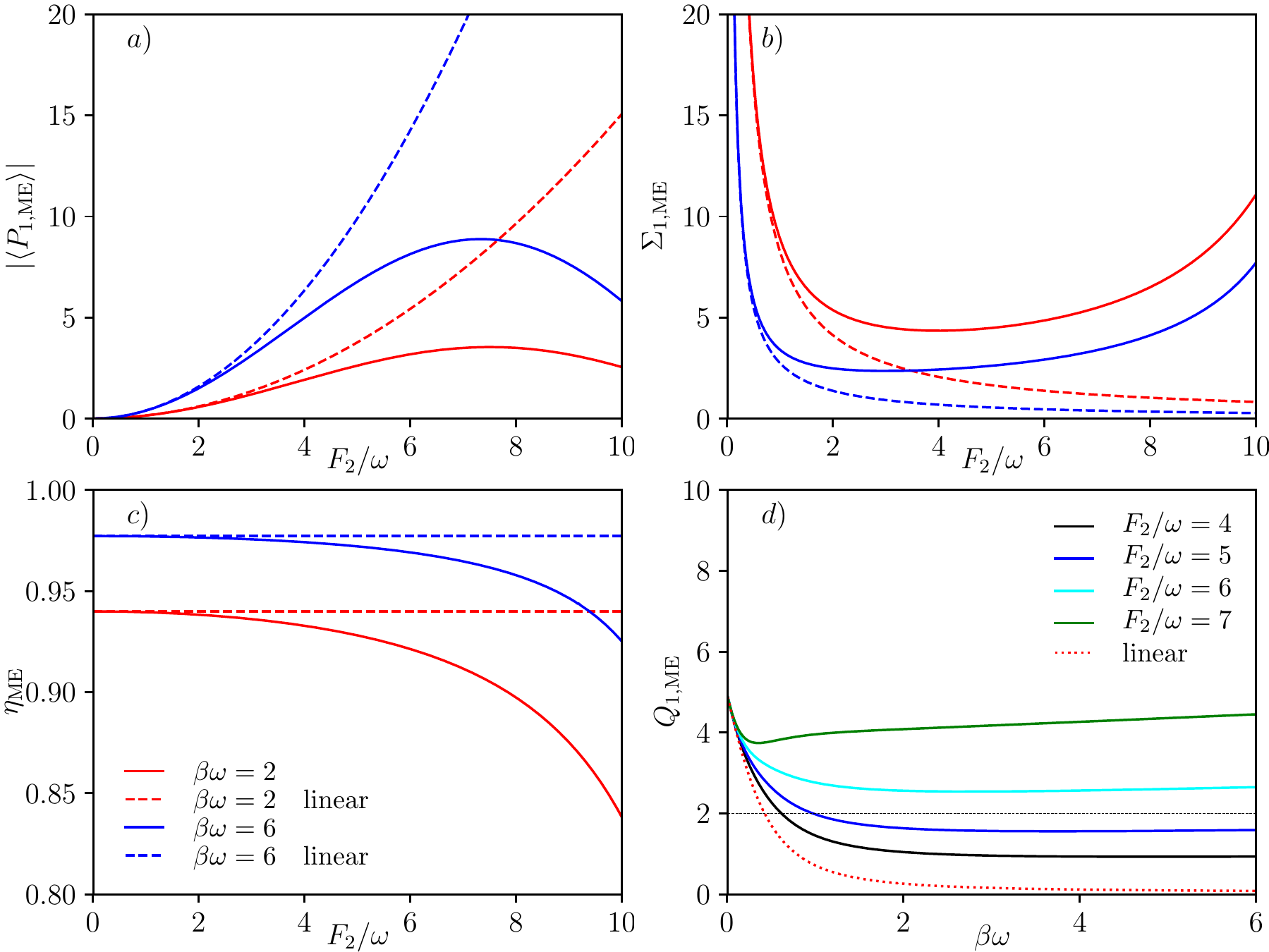}
\caption{ Non-linear regime. Panel a) and b): mean power and power fluctuations  at maximum efficiency versus  $F_2/\omega$. Panel c): efficiency $\eta_{\rm ME}$ versus $F_2/\omega$.
 Panel d): Trade-off criterion $Q_{1,\rm ME}$ versus $\beta\omega$ for different $F_2/\omega$. See text.
The solid curves are those of nonlinear response (NLR), and the dashed curves pertain to linear response (LR). The
parameters are $K=0.1$,  $\alpha=5$, $\omega=5\Delta_{{\rm r}}$, and $\Delta_{{\rm r}}=1$.
The parameter range covered by the plots is experimentally accessible in QPC transport experiments 
in the FQH regime~\cite{dubois2013,glattli}}. 
\label{fig:three}
\end{figure}

The characteristic behaviors of  $\langle P_{1,\rm ME}\rangle $, $\eta_{1,\rm ME}^{}$, and $\Sigma_{1,\rm ME}$  versus
$F_2/\omega$ in the NLR are shown in  Fig.~3 (a)-(c) for $\beta\omega=6$ (blue) and
$\beta\omega=2$ (red) for $K=0.1$. Clear deviations from the LR behaviors occur
in (a), (b), and (c), as $F_2/\omega$ is increased.
The yield $|\langle P_{1,\rm ME}\rangle|$  reaches a maximum near $F_2/\omega=7.5$ for both temperatures.
By contrast, the NLR power fluctuations run through a flat minimum located near $F_2/\omega =2$ for $\beta\omega=6$ and near $F_2/\omega=4$ for $\beta\omega=2$ and spanning a broad amplitude range.
In this area,  the work-to-work converter has sizeable power yield with simultaneous low power fluctuations and efficiency still close to unity, only slightly smaller than in LR.
This indicates that the NLR regime is a promising field for finding best compromise between large power yield, low power fluctuations and high efficiency.
Panel (d) displays the trade-off criterion $Q_{1,\rm ME}$ versus $\beta\omega$
for different values of $F_2/\omega$. Most interestingly, the quantity $Q_{1,\rm ME}$ falls below 2 for $F_2/\omega$ below 5.5 and sufficiently low temperature. On the contrary, it consistently stays above 2 for larger $F_2/\omega$ and arbitrarily low temperatures. In the former case,  the power fluctuations are in the flat minimum
of panel (b), thereby facilitating shortfall of the trade-off bound in the NLR.

So far, we have studied mean powers and power dispersion  of the QBP converter in order~$\Delta^2$. Contributions of higher order in  $\Delta^2$
may become significant at sufficiently low $T$, depending on the parameters of the model. Starting out from the
real-time version~\cite{weissbook} of the Coulomb gas representation~\cite{yuval70} of the perturbative series in $\Delta^2$, we have calculated the  $\Delta^4$ terms of powers and variance numerically. These terms result from direct next-to-nearest-neighbor transitions in the TB lattice and coherent transport of two charges in the associated fermionic transport model. In addition to this, we have approximately taken into account all tunneling
contributions of higher order of $\Delta^2$ by summation of partial contributions in each order. The quality of this
approximate treatment of the strong tunneling regime has been checked
for the  point $K=\frac{1}{2}$, for which all orders in $\Delta^2$  can be summed
exactly~\cite{sassetti92,weissbook}. Formidable agreement  down to very low temperatures has been found.
The conclusions of the numerical analysis are that the higher-order tunneling terms yield marginal contributions  up to inverse temperature $\beta\omega=6$ for $\Delta_{\rm r}=1$, and the above weak-tunneling results are qualitatively correct down to much lower temperatures. Until now, reliable results in the asymptotic low temperature regime are missing. Nevertheless, it is rather unlikely that coherent tunneling transitions across many TB states will spoil the characteristics shown above.

{\it Conclusions ---}
We have studied work-to-work conversion of a quantum Brownian particle in a TB lattice subjected to two isochromatic drives and coupled to a thermal bath with Ohmic spectral density. {We have argued that  this scenario can be experimentally realized and tested by a two-terminal setup of a fractional quantum Hall bar with a quantum point contact.     }
Analytic results in the linear response regime  have been presented for mean power, efficiency, power fluctuations,
and the trade-off  criterion.
It has been  shown that optimal performance at weak damping and low temperatures comes along with
a clear undercut of the classical trade-off bound.
We have also focussed on the performance in the regime of nonlinear response to driving with large amplitudes.
It has been found that large power yield with low power fluctuations and with efficiency close to unity can be realized in a wide parameter range of the external drive. This uncloses the hitherto mostly unregarded nonlinear response regime
as a promising new operation field for isothermal machines.
\begin{acknowledgments}
We wish to thank Udo Seifert for stimulating discussions. M.C. acknowledges support from the project Quant-Eranet ``SuperTop". M. S. thanks UniGE for financial support.
\end{acknowledgments}

\appendix
\newpage
\setcounter{equation}{0}

\section{Supplemental Material}
\centerline{\bf Power and power fluctuations}
\vspace{1mm}
The Hamiltonian of the quantum Brownian particle (QBP) in a TB lattice with lattice spacing $q_0$, bilinearly coupled to a bath of harmonic oscillators and driven by two time-periodic forces of period ${\cal T}$, $\hbar\epsilon_1(t)/q_0$ and  $\hbar\epsilon_2(t)/q_0$, 
is~\cite{sassetti92,weissbook}
\be\label{s1}
H(t) = H_{\rm TB} + H_{\rm RI} + H_{\rm ext}(t) \, ,
\ee
where
\beq \nonumber
H_{\rm TB} &=& -\frac{\hbar\Delta}{2}\sum_n(a_n^\dagger a_{n+1}^{}+{\rm h.c.}) \, ,  \\ \label{s2}
 H_{\rm RI} &=& \sum_\alpha[\frac{p_\alpha^2}{2m_\alpha}+ \frac{m_\alpha\omega_\alpha^2}{2}(x_\alpha-\frac{c_\alpha}{m_\alpha\omega_\alpha}q)^2] \, ,  \\  \nonumber
H_{\rm ext}(t) &=& - \hbar [\epsilon_1(t)+\epsilon_2(t)]/q_0 \, .
\eeq
Here $q=q_0\sum_na_n^\dagger a_n^{}$, and $\hbar\Delta$ is the tunneling coupling energy of neighboring TB states.
The spectral density of the bath coupling is
$J(\omega) = \frac{\pi}{2}\sum_\alpha
\frac{c_\alpha^2}{m_\alpha\omega_\alpha}\delta(\omega-\omega_\alpha)$.
From now on we put $q_0=\hbar=k_{{\rm B}}=1$.

The power $P_i(t)$ ($i=1,2$) for the drive $\epsilon_i^{}(t)$ is related to the Brownian particle's
velocity $\dot{q}(t)$ as
\be  \label{s3}
P_i(t)=\epsilon_i(t)\dot{q}(t)\,.
\ee

The TB representation of the average position of the Brownian particle is a perturbative series in $\Delta^2$.  It can be written as a grand-canonical sum of a 1D gas of charges $e_j^{}=\pm 1$with complex interactions 
${\rm e}^{e_i^{} e_j^{}[\, W'(\tau_{ij}^{})\pm i\,  W''(\tau_{ij}^{})\,]  }$, where $\tau_{ij}^{}$ is the 
distance of the charge pair ~\cite{sassetti92,weissbook,carrega15}.  
The complex pair interaction $W(\tau)=W'(\tau) +i\,W''(\tau)$ includes all effects of the spectral bath coupling, and is defined as
\be \label{s4}
W(\tau)=\frac{1}{\pi}\int_0^{\infty}d\omega \frac{J(\omega)}{\omega^2}\frac{\cosh\big[\omega{\frac{\beta}{2}}\big]-\cosh\big[{\omega(\frac{\beta}{2}-i\tau)}\big]}{\sinh\big[{\omega\frac{\beta}{2}}\big]}  \, .
\ee
For complex time $z=t-i\,\tau$, the equilibrium  correlation function $W(z)$ is analytic in the strip $0\ge{\rm Im}\,z>-\beta$ and satisfies
\be\label{sdb}
W(-z-i\,\beta) = W(z) \, .
\ee

In the weak tunneling limit, the position of the quantum Brownian particle at time $t$ is
\be
q(t)=\int_0^t dt_2\int_0^{t_2}dt_1~k_{\rm P}(t_2-t_1)\sin[G(t_2,t_1)]\, ,
\label{s5}
\ee
where
\be
\label{kPsm}
k_{\rm P}(\tau)=\Delta^2\sin[W''(\tau)]e^{-W'(\tau)} 
\ee
includes the bath correlations, and $G(t_2,t_1)$ is the total bias phase accumulated in the time interval extending from 
$t_1$ to $t_2$,
\be\label{s7}
G(t_2,t_1)= \sum_{i=1,2}\int_{t_1}^{t_2}dt'\,\epsilon_i(t') \, .
\ee
At times much larger than the  decay time of $k_{\rm P}(\tau)$ (indicated by the overbar), the
power   $\overline{P}_i(t)= \epsilon_i(t)  \overline{\dot{q}}(t)$, $ i=1,\, 2,$ is 
\be  \label{s11}
\overline{P}_i(t)=\epsilon_i(t) \int_0^{\infty}d\tau~k_{\rm P}(\tau)\sin[G(t,t-\tau)]\, .
\ee
The function $\overline{P}_i(t) $ is a periodic function of $t$ with period ${\cal T}$. The steady-state component 
$\langle P_i\rangle$ of $\overline{P}_i(t) $ is obtained upon taking the average over the period ${\cal T}$.  
\be
\label{S12}
\langle P_i\rangle= \int_0^{\infty}d\tau\,k_{\rm P}(\tau)\int_0^{\cal T}\frac{d t}{{\cal T}}\,
\epsilon_i(t)\sin[G(t,t-\tau)]  \, .
\ee

Consider next the power variance, which is defined as 
\be
D_i(t)=\frac{1}{t}\int_0^t dt_2\int_0^t dt_1~\delta P_i(t_2)\delta P_i(t_1)~ ~,
\label{S9}
\ee
where $\delta P_i (\tau)= P_i(\tau) - \overline{P_i}(\tau)$. At long times, we then have
\be
\overline{ D_i}(t)=\int_0^\infty d\tau\,
[\,\overline{\delta P_i(t)\delta P_i(t-\tau)}+ \overline{\delta P_i(t-\tau)\delta P_i(t)}\, ]   \, ,
\ee
and with the relation $ P_i(t)  = \epsilon_i(t)\dot{q}(t)$
\begin{eqnarray} \label{S11}
\overline{D_i}(t)&=& \int_0^{\infty}d\tau \; \epsilon_i(t)\epsilon_i(t-\tau)    \\    \nonumber
&\times&\big[\, \overline{\dot{q}(t)\dot{q}(t-\tau)}+\overline{\dot{q}(t-\tau)\dot{q}(t)} - 
2\overline{\dot{q}}(t)\overline{\dot{q}}(t-\tau) \,\big]\,.
\end{eqnarray}
From this, the steady-state component is found in order $\Delta^2$ as
\be \label{S14}
\langle { D}_{i}\rangle= \!\!\! \int_0^{\infty}\!\! d\tau\,k_{\rm D}(\tau)\!\!\int_0^{\cal T}\!\frac{d t}{{\cal T}}~\epsilon_i(t)\epsilon_i(t-\tau)\cos[G(t,t-\tau)] \, ,
\ee
where
\be   \label{kDsm}
k_{\rm D}(\tau) = \Delta^2\,\cos[W''(\tau)]\, e^{-W'(\tau)}  \, .
\ee
The property (\ref{sdb}) leads to the detailed balance relation~\cite{weissbook} 
\be \label{dbr}
\int_0^\infty\!\!\! d\tau\,\sin(\omega\tau) k_{\rm P}(\tau) = 
\coth\Big(\frac{\omega\beta}{2}\Big)\int_0^\infty\!\!\! d\tau\,\cos(\omega\tau) k_{\rm D}(\tau)  \, .
\ee
The Ohmic spectral density of the coupling is $J(\omega) = 2\pi K\,\omega$, where $K$ is the dimensionless coupling strength. 
Upon including modes above a cut-off frequency $\omega_{\rm c}$ in adiabatic approximation, we obtain
from Eq.~(\ref{s4})  the analytic form
\be \label{s8}
W(\tau)=2K\ln\left[(\beta\omega_c/\pi)\sinh{(\pi|\tau|/\beta)}\right] +i\pi K{\rm sgn}(\tau)\,.
\ee
With the dressed tunneling amplitude 
$\Delta_{\rm r} = \Delta (\Delta/\omega_{\rm c})^{K/(1-K)}$,  
the functions (\ref{kPsm}) and (\ref{kDsm}) take in the range $\tau>0$  the forms
\beq  \nonumber
k_{\rm P}(\tau) &=& \Delta_{\rm r}^2 \Big(\frac{\pi}{\beta \Delta_{\rm r}}\Big)^{2 K}
\frac{\sin(\pi K)}{\sinh[\pi\tau/\beta]^{2 K}}\, , \\       \label{s9}
k_{\rm D}(\tau) &=& \Delta_{\rm r}^2 \Big(\frac{\pi}{\beta \Delta_{\rm r}}\Big)^{2 K}
\frac{\cos(\pi K)}{\sinh[\pi\tau/\beta]^{2 K}}\, .
\eeq
\vspace{3mm}

\centerline{\bf Brownian duet}
\vspace{1mm}
Consider mean power and power variance for isochromatic driving 
with an added multiple frequency term in the output channel, and a phase shift
$\varphi = \arctan\alpha$ in the input channel,
\beq  \nonumber
\epsilon(t) &=& \epsilon_1(t)+\epsilon_2(t) \, , \\     \label{sdrive}
\epsilon_1(t) &=& F_1\,[ \sin(\omega t)+ \gamma_n \sin(n\omega t)]\, , \\  \nonumber
\epsilon_2 (t) &=& F_2\cos(\omega t - \varphi)\, .
\eeq
With the drive (\ref{sdrive}), the bias phase can be written as
\beq \nonumber
G(t,t-\tau) &=&\frac{ {\cal F}}{\omega}\big\{\cos[\omega(t-\tau)+\Phi]-\cos[\omega t+ \Phi]\big\} \\
&+& \frac{F_1}{n\omega} \gamma_n \big\{  \cos[n\omega (t-\tau)]- \cos(n \omega t)\big\}  \, ,
\eeq
where ${\cal F} ={\rm sgn}\,(F_1+F_2 \sin\varphi) [F_1^2+F_2^2+2F_1 F_2\sin\varphi]^{1/2}$, and 
$\Phi = \arctan[F_2 \cos\varphi/(F_1+F_2 \sin\varphi)] $. The bias phase factor
\be
B(t,t-\tau)=e^{i \, G(t,\,t-\tau)} 
\label{eq:sppsm}
\ee
is a periodic function of time $t$. It can be written as the  double Fourier series
\beq  \nonumber
B(t,t-\tau)&=&\sum_{k=-\infty}^{+\infty}  J_k [A(\tau)]\,e^{-i\, k \omega\tau/2}e^{i\, k \Phi} \\  \label{eq:sb}
    &&\times \sum_{\ell=-\infty}^{+\infty}  J_\ell [A_1(\tau)]\,e^{-i\, \ell n\omega\tau/2}
\, e^{i \, (k + \ell n) \omega\, t}  \, ,
\eeq
where $J_k(z)$ is a Bessel function, and where
\beq  \nonumber
A(\tau) &=& 2{\cal F} \sin(\omega\tau/2)/\omega \, , \\
A_1(\tau) &=& 2 F_1 \gamma_n \sin(n\omega\tau/2)/(n\omega) \, .
\eeq
With the  Fourier series (\ref{eq:sb}), the time averages in Eqs.~(\ref{S12}) and (\ref{S14})
are straightforward, yielding for $j=1,\,2$
\beq\nonumber
\langle P_j\rangle &=&  \int_0^{\infty} \! d\tau~k_{\rm P}(\tau) p_j(\tau) \, ,    \\   \label{eq:PDjmean}  
\langle D_j\rangle &=&  \int_0^{\infty} \! d\tau~k_{\rm D}(\tau) d_j(\tau)\, .    
\eeq
The functions $p_j(\tau)$ and $d_j(\tau)$ are defined by single infinite sums, in which the coefficients are products of two 
$J$-Bessel functions times phase factors. For $\gamma_n=0$, we have $A_1(\tau)=0$. 
Then the sums reduce to individual contributions. These are
\beq  \nonumber 
 p_1(\tau) &=& F_1  J_1[A(\tau)]\cos(\Phi-\omega\tau/2)\, ,  \\    \label{eq:p12}
p_2(\tau) &=&F_2   J_1[A(\tau)] \,\sin(\Phi+\varphi-\omega\tau/2) \, .
\eeq
and 
\beq
\label{eq:d12}
d_{1}(\tau) &=&  \frac{F_1^2}{2}  \big[J_0[A(\tau)]\cos(\omega\tau) - J_2[A(\tau)]\cos(2\Phi) \big] \, ,   \\
\nonumber
d_{2}(\tau) &=&\frac{F_2^2}{2} \big[ J_0[A(\tau)]\cos(\omega\tau) + J_2[A(\tau)]\cos(2\Phi+2\varphi) \big] .
\eeq
The expressions (\ref{eq:PDjmean}) with (\ref{eq:p12}) and  (\ref{eq:d12}) yield the expressions (8) and (9)  of the Letter.
\vspace{5mm}

\centerline{\bf Linear response}
\vspace{1mm}

\noindent
In linear thermodynamics, the fluxes are linear in the  forces, and the powers are
quadratic forms of the forces, $\langle P_i\rangle =F_i \sum_{j=1,2} {\cal L}_{i,j}\, F_j$, where ${\cal L}$
is the  Onsager matrix. Expanding the general expression (\ref{s11})  up to terms quadratic in the forces 
$F_1$ and $F_2$, and taking the time average, we get
\be  \label{Slin1}
{\langle P_i\rangle} = \int_0^{\infty}d\tau~k_{\rm P}(\tau)\int_0^{\cal T}\frac{d t}{{\cal T}}~\epsilon_i(t) G(t,t-\tau) \, .
\ee
From this, the Onsager matrix can be extracted as
\be \label{Slin2}
{\cal L}_{i,j}=\int_0^{\infty}d\tau~k_{\rm P}(\tau)\int_0^{\cal T}\frac{d t}{{\cal T}}~\frac{\epsilon_i(t)}{F_i} \int_{t-\tau}^{t}dt'~\frac{\epsilon_j(t')}{F_j} \, .
\ee
For the drive (\ref{sdrive}), we get
\beq \label{eq:onsmatrixSM}
{\cal L}= L_{\,\rm s}(\omega) \left(\begin{array}{cc}
1+\rho_n & \frac{\alpha + \kappa}{\sqrt{1+\alpha^2}} \\
\frac{\alpha -\kappa}{\sqrt{1+\alpha^2}} & 1
\end{array}\right) \, ,
\eeq
where $\kappa = L_{\rm c}(\omega)/L_{\rm s}(\omega)$, and 
$\rho_n = \gamma_n^2 L_{\rm s}(n\omega)/L_{\rm s}(\omega)$.
The functions $L_{\,\rm s}(\omega)$ and  $L_{\,\rm c}(\omega)$ are 
\beq  \label{ls}
 L_{\,\rm s}(\omega)&=&\int_0^{\infty}d\tau~ k_{\rm P}(\tau)\frac{\sin(\omega \tau)}{2\omega}  \\ \label{lc}
  L_{\,\rm c}(\omega) &=&\int_0^{\infty}d\tau~k_{\rm P}(\tau)\frac{1-\cos(\omega\tau )}{2\omega}~.
\eeq

The power variance in steady-state  $\langle D_1 \rangle$ to second order in the force is given by
\beq  \nonumber
\langle { D}_{1}\rangle &=&  \int_0^{\infty} d\tau\,k_{\rm D}(\tau)\int_0^{\cal T}\!\!\!
\frac{d t}{{\cal T}}~\epsilon_1(t)\epsilon_1(t-\tau)     \\ \label{di}
&=& \frac{F_1^2}{2} \int_0^\infty d\tau \, k_{\rm D}(\tau)\,
[\cos(\omega \tau) + \gamma_n^2\cos(n\omega \tau)]  \, .
\eeq
Observing the detailed balance relation (\ref{dbr}), we finally obtain
\be
\langle { D}_{1}\rangle=F_1^2 \omega\coth(\beta\omega/2)\, \sigma_n L _{\,\rm s}(\omega) \, ,
\ee
where
\be  \label{sign}
\sigma_n = 1 +\gamma_n^2 n \,
 \frac{ \coth(\beta n \omega/2) L_{\rm s}(n\omega)}{\coth(\beta\omega/2) L_{\rm s}(\omega)}  \, .
\ee

With the function (\ref{s8}), the integrals (\ref{ls}) and (\ref{lc}) can be calculated in analytic form.
The resulting expressions are  given in terms of Euler's function $\Gamma(z)$ as
\beq \label{Lsana}
 L_{\,\rm s} (\omega)&=& \frac{1}{2\beta\omega}\Big( \frac{\beta \Delta_{\rm r}}{2\pi} \Big)^{2-2K} 
\sin(2\pi K)\Gamma(1-2 K)  \\   \nonumber
&&\quad\times\; \textstyle{ \sinh\Big(\frac{\beta\omega}{2}\Big)
\Gamma\Big( K-i\,\frac{\beta\omega}{2\pi} \Big)
\Gamma\Big( K+i\, \frac{\beta\omega}{2\pi}\Big) }   \,,    \\  \label{Lcana}
 L_{\,\rm c}(\omega) &=&
 \frac{1}{\beta\omega}\Big( \frac{\beta \Delta_{\rm r}}{2\pi} \Big)^{2-2K} 
\sin(\pi K)^2 \Gamma(1-2 K)    \\  \nonumber
&\times&\Big[\Gamma(K)^2-\Gamma\Big(K-i\,\textstyle{\frac{\beta\omega}{2\pi} \Big)\Gamma\Big(K+i\,\frac{\beta\omega}{2\pi} \Big)\cosh\Big(\frac{\beta\omega}{2} }\Big)\Big]~.
\eeq
The ratio of these functions, $\kappa = L_{\,\rm c}(\omega)/L_{\,\rm s}(\omega)$, is
\beq \label{kappaana}
\kappa &=& \frac{\tan(\pi K)}{\sinh(\frac{\beta\omega}{2})}  \\   \nonumber
&&\times\; \bigg[\,   \frac{\Gamma(K)^2}{  \Gamma\big( K-i\,\frac{\beta\omega}{2\pi} \big)
\Gamma\big( K+i\, \frac{\beta\omega}{2\pi}\big)}   - \textstyle{\cosh\big(\frac{\beta\omega}{2}\big)} \,  \bigg] \, .
\eeq
\vspace{0mm}

\noindent
{\it Maximum output power. ---}\\[2mm]
For the drive (\ref{sdrive}), the maximum output power or yield is at
\be
F_1 =F_{1,\rm MP} \equiv - \frac{\alpha+\kappa}{2\sqrt{1+\alpha^2}}\frac{F_2}{1+\rho_n} \; ,
\ee
yielding
\be
\langle P_{1,\rm MP}\rangle = - (1+\rho_n) L_{\rm s}(\omega) F_{1,\rm MP}^2 \, .
\ee
The two branches for the efficiency $\eta=|\langle P_1\rangle|/\langle P_2\rangle$ and the relative uncertainty 
$\Sigma =\sqrt{\langle  D_1\rangle/\langle P_1\rangle^2}$ as functions of $P^\ast$, resulting from the condition
$\langle P_1(F_1,F_2)\rangle/\langle P_{1,\rm MP}\rangle= P^\ast$, are
\beq
\label{eq:etabranchessm}
\eta_{\pm}^{} &=&\frac{P^*}{2}\frac{X}{1+2/Y\mp\sqrt{1-P^*}} \, , \\   \nonumber
\Sigma_{1,\pm}&=& \frac{2}{F_2}
 \frac{\sqrt{1+\alpha^2}}{\alpha+\kappa}\sqrt{\frac{\sigma_n \omega\coth(\beta\omega/2)}{L_{\,\rm s}(\omega)} }
\frac{1\pm\sqrt{1-P^*}}{P^* }  \, ,
\eeq
where 
\beq\nonumber
X &=&\alpha + \kappa)/(\alpha-\kappa)\, , \\
Y&=&(\alpha^2-\kappa^2)/ (1+r_n+ \kappa^2)  \, ,   \\   \nonumber
r_n &=& \gamma_n^2 (1+\alpha^2) L_{\rm s}(n\omega)/ L_{\rm s}(\omega)  \,  .
\eeq
In the absence of the multiple frequency drive, $\gamma_n=0$, these forms reduce to the expressions (13) and (14)
of the Letter.
\vspace{4mm}

\noindent
{\it Maximum efficiency. ---}\\[2mm]
The efficiency $\eta$ is maximal at $F_1=F_{1,\rm ME}$, where
\be  \label{F1ME}
F_{1,\rm ME} =- \frac{\sqrt{1+\alpha^2}}{\alpha-\kappa}
\left( 1 - \sqrt{ \frac{1+r_n+\alpha^2}{1+r_n+\kappa^2} } \right) F_2  \, .
\ee 
It is given by
\be
\label{etaME}
\eta_{\rm ME}^{}= \frac{\sqrt{1+r_n+\alpha^2}-
\sqrt{1+r_n+\kappa^2}}{\sqrt{1+r_n+\alpha^2}+\sqrt{1+r_n+\kappa^2}}\;
\frac{\alpha +\kappa}{\alpha-\kappa}~\, .
\ee
The corresponding mean power and power fluctuations are
\beq
\label{P1ME}
\langle P_{1,\rm ME}\rangle &=& - F_2^2 \eta_{\rm ME}\, L_{\,\rm s}(\omega)
\sqrt{\frac{(1+r+\kappa^2)}{(1+\alpha^2)(1+\rho_n)} }\, ,\nonumber \\
\Sigma_{1,\rm ME} &=&\sqrt{\sigma_n  L_{\,\rm s}(\omega) \,\omega\coth(\beta\omega/2)} \;
F_{1,\rm ME}/\langle P_{1,\rm ME}\rangle \, .
\eeq

With the expressions (\ref{etaME}) and (\ref{P1ME}), the trade-off quantity 
\be \label{Q1MEdef}
Q_{1,\rm ME} = \beta |\langle P_{1,\rm ME}\rangle|\,(1/\eta_{\rm ME}-1)\, \Sigma_{1,\rm ME}^2 \,
\ee
is found in analytic form as
\beq  \nonumber
Q_{1,\rm ME} &=&c(\beta\omega) \frac{\sigma_n}{\sqrt{1+\rho_n}}\frac{\sqrt{1+\alpha^2}}{(\alpha+\kappa)^2 } 
\\   \label{Q1ME} 
&& \times \left( \frac{1+r_n-\alpha\kappa}{\sqrt{1+r_n+\kappa^2}}+\sqrt{1+r_n+\alpha^2}\right)  \, ,
\eeq
where $c(x)= 2x \coth(x/2) $. In the absence of the multiple frequency term in the  drive $\epsilon_1(t)$, $\gamma_n=0$,
we have $r_n=0$ and $\sigma_n=1$, and thus the expressions (\ref{etaME}) and (\ref{Q1ME}) reduce to the expression (15)
and (17) of the Letter.

In the asymptotic low temperature  regime $\beta\omega\gg1$, we obtain from (\ref{Lsana}) and (\ref{kappaana})
\beq  \nonumber
L_{\rm s}^{\rm (as)}(\omega) &=& \frac{1}{4} \sin(2\pi K)\Gamma(1-2K)\bigg(\frac{\Delta_{\rm r}}{\omega}\bigg)^{2-2K}
  \, ,\\
\kappa_{}^{\rm (as)} &=& \tan(\pi K) \bigg[    \frac{\Gamma(K)^2}{\pi}\bigg( \frac{\beta\omega}{2\pi} \bigg)^{1-2K}   - 1 \bigg]\, ,
\eeq
and
\beq     \nonumber
\rho_n^{\rm (as)} &=& \gamma_n^2/n^{2-2K} \, , \\ \nonumber
\sigma_n^{\rm (as)} &=& 1 + \gamma_n^2/n^{1-2K} \, , \\   \label{rhoas}
r_n^{\rm (as)} &=& (1+\alpha^2) \gamma_n^2/n^{2-2K} \, .
\eeq
Since $\kappa_{}^{\rm (as)}$ diverges for $K<1/2$, as $\beta\to \infty$, whereas $\rho_n^{\rm (as)}$, 
$\sigma_n^{\rm (as)}$ and $r_n^{\rm (as)} $ are temperature-independent in this limit,  the qualitative behaviors of 
$\eta_{\rm ME}$ and $Q_{1,\rm ME}$ are independent of the coupling parameter $\gamma_n$. 
Altogether, the behaviors of  efficiency, power fluctuations and trade-off $Q_{1,\rm ME}$
displayed in Figs.~1 and 2  of the Letter change only marginally, when a multiple frequency contribution is added to the base-frequency term in the output $\epsilon_1(t)$.

When the multiple frequency contribution is added in the output channel, as in Eq.~(\ref{sdrive}), 
the element ${\cal L}_{1,1}$ of the Onsager matrix is modified by a factor $1+\rho_n$.
If instead we had added a multiple frequency term in the input channel $\epsilon_2(t)$, the element  ${\cal L}_{2,2}$ of the Onsager matrix, would be changed by a factor $1+\rho_n$.   With the argumentation similar to that below 
Eq.~(\ref{rhoas}), one would find again that  efficiency, power fluctuations, and trade-off $Q_{1,\rm ME}$ change only marginally, when a multiple frequency contribution is added  to the base-frequency term in the input $\epsilon_2(t)$.  

\newpage
\begin{center}
{\bf Mapping with quasiparticle tunneling through \\a quantum point contact (QPC) in a FQH system}
\end{center}
\vspace{1mm}
Consider a fractional quantum Hall (FQH) bar (see Fig.~1) with Laughlin filling factor $\nu =1/(2n+1)$ $(n \in \mathbb N$) described in hydrodynamical formulation \cite{wen90} by the model Hamiltonian 
\be \label{hamtot}
H = H_0 + H_{\rm g} + H_{\rm T}  \, ,
\ee
\beq  \nonumber
H_{ 0} &=& \frac{\upsilon}{4\pi } \int_{-\infty}^{+\infty} dx
 (\left[ \partial_x \phi_{\rm R}^{}(x) \right]^2 +\left[ \partial_x \Phi_{\rm L}^{}(x) \right]^2 ) \, ,  \\   \nonumber
H_{\rm g } &=& -e \int_{-\infty}^{\infty}\!\! dx \, \big[ \Theta(- x-d) V_1(t) \rho_{\rm R}^{}(x) \\   \nonumber
  &&\qquad \qquad \;\;+\; \Theta(x-d) V_2(t)\rho_{\rm L}^{}(x) \big]  \, ,    \\  \label{hamsingle}
H_{\rm T}&=&\Gamma_0 \;\big[\Psi_{\rm R}^{\dagger}(0)\Psi_{\rm L}^{}(0)+\text{h.c.} \big] \, .
\eeq
Here, $H_0$ describes  the chiral edge states  with propagation direction ${\rm R/L}$ and velocity $\upsilon$ 
in terms of bosonic fields $\phi_{\rm R/L}$. The term $H_{\rm g}$ represents capacitive coupling of the densities 
$\rho_{\rm R/L}^{}(x)= \mp \frac{\nu}{2\pi}\partial_x \phi_{\rm R/L}(x)$ with two voltage gates acting separately on the right and left moving excitations. The step functions $\Theta(\mp x-d)$ describes the case of 
very long contacts, which is in accordance with standard experimental setups \cite{dubois2013, glattli}. The contacts are separated by distance $2 d$. Weak backscattering transfer of quasiparticles between the two edges at the QPC located at $x=0$ is described by the tunneling term $H_{\rm T}$. Here, 
$\Psi_{\rm R/L}(x) =(\,U_{\rm R/L}/\sqrt{2\pi a}\,)\, {\rm e}^{\pm i\,k_{\rm F} x}\,
{\rm e}^{-i\,\sqrt{\nu} \phi_{\rm R/L}(x)}_{}$ is the quasiparticle annihilation operator, $a$ is a cut-off length, 
$U$ a Klein factor, and $k_{\rm F}$ is the Fermi momentum.
\begin{figure}
\centering
\includegraphics[width=0.7\columnwidth]{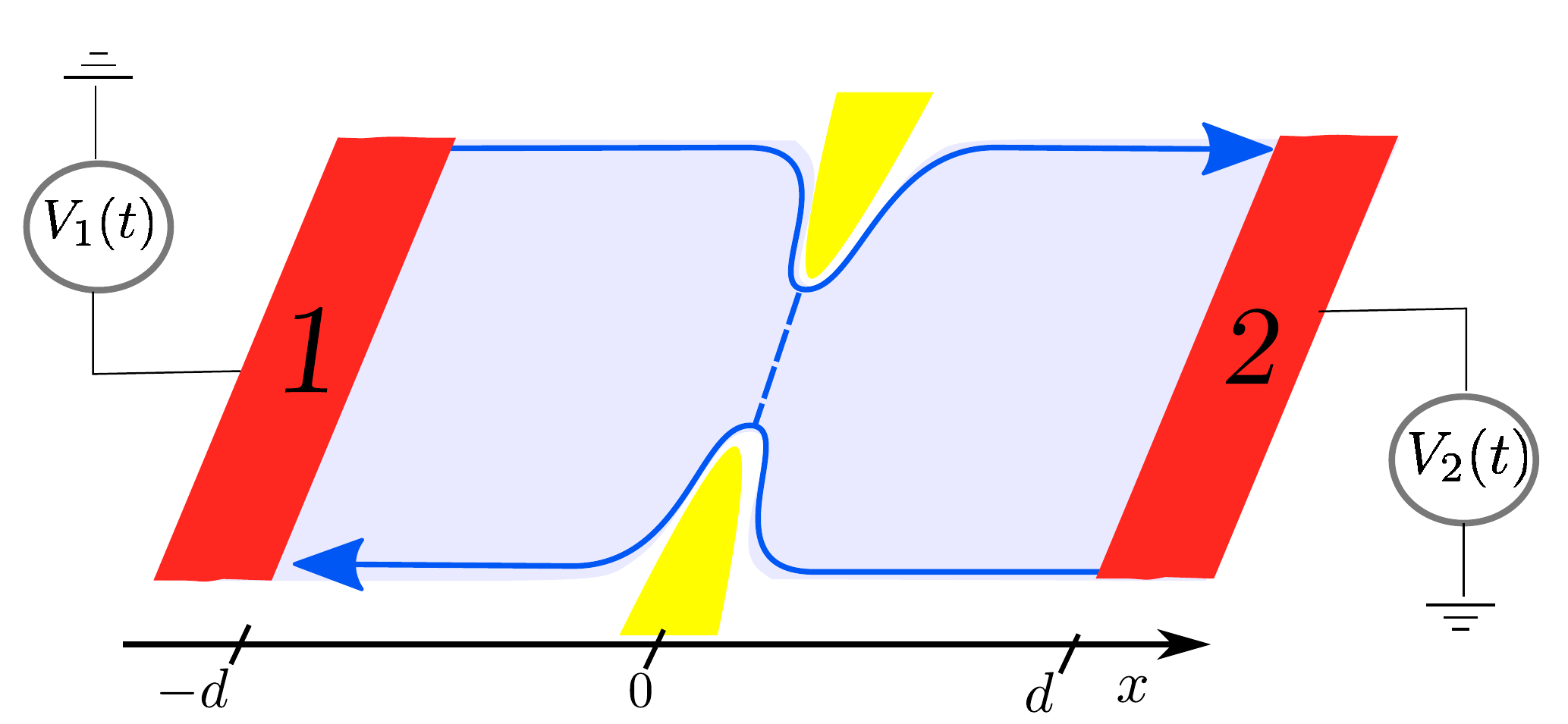}
\caption{Sketch of the two terminal setup  of a fractional quantum Hall bar with a quantum point contact (QPC). The QPC is placed at $x=0$, and the two time-dependent voltages are applied at $x=-d$ and $x=d$.}
\label{setup}
\end{figure}

In the absence of the QPC, the currents at the terminals placed at $x=\pm d$ are the right/left 
moving edge currents,
\begin{align}
 J_{x=\pm d}(t) &= J_{0,\rm R/L}(t) = G_0 \, V_{1/2}\left(t- 2d/\upsilon\right)  \, ,  \label{eq:j0_av}
\end{align}
where $G_0= \frac{\nu e^2 }{2\pi}$ is the universal quantum of conductance in the FQH regime.
In the presence of the QPC, these current are modified by the backscattering current $J_{\rm bs}(t)$  of the quasiparticles as
\begin{align}  \nonumber
&J_{x=-d}(t)= J_{0,\rm L}(t) + J_{\rm bs}(t) \, ,\\ \label{Jterms}
&J_{x=d}(t)= J_{0,\rm R}(t) - J_{\rm bs}(t) \, , 
\end{align}
and the associated powers are
\begin{align}  \nonumber
&P_{x=-d}(t)= V_1(t)[ J_{0,\rm L}(t) + J_{\rm bs}(t)] \, ,\\  \label{Pterms}
&P_{x=d}(t)= V_2(t) [J_{0,\rm R}(t) - J_{\rm bs}(t)]  \, .
\end{align} 
The powers $P_{1/2}(t)$ resulting from backscattering alone are
\begin{align}  \nonumber
&P_1(t) \equiv  P_{x=-d}(t)- V_1(t) J_{0,\rm L}(t) =  V_1(t)  J_{\rm bs}(t) \, ,\\  \label{Pexcess}
&P_2(t) \equiv P_{x=d}(t)- V_2(t) J_{0,\rm R}(t)  =- V_2(t)  J_{\rm bs}(t)  \, .
\end{align}

Following the analysis set out in Refs.~\cite{vannucciprl,ronettiprb}, the backscattering current for weak quasiparticle
tunneling is found as
\beq  \nonumber
J_{\rm bs}(t) &=& 2i\nu e  \Big(\frac{\Gamma_0}{2\pi a} \Big)^2  \int_{0}^{\infty} \!\!\!d\tau 
\sin\Big[\nu e \int_{t-\tau}^{t}\!\! dt'\,V_{-}(t')\Big]\\ \label{bsc}
&& \times\left(e^{2\nu\,{\cal G}(\tau)}- e^{2\nu\,{\cal G}(-\tau)}\right) \, ,
\eeq
where $V_{-}(t)=V_{1}(t)-V_{2}(t)$, and ${\cal G}(\tau)$  is the connected Green's function of the 
quasiparticle field $\phi(x,\tau)$ at $x=0$, 
${\cal G}(\tau) =\langle \phi_{\rm R/L}(0,\tau)  \phi_{\rm R/L}(0,0)\rangle_{\rm c} $.
Upon equating the filling factor $\nu$ with the Ohmic damping parameter $K$, and the length $a$ with 
$\upsilon/\omega_{\rm c}$, there directly holds in the scaling limit the correspondence
\be
 2\nu {\cal G}(\tau) = - W(\tau)  \, ,  \\[1mm]
\ee
where $W(\tau)$ is the Ohmic bath correlation function (\ref{s8}). 
With the correspondences $\epsilon_1(t) = \nu e\, V_1(t)$, $\epsilon_2(t) = -\nu e\, V_2(t)$,
$G(t_2,t_1) = \nu e\int_{t_1}^{t_2} dt'\,[V_1(t')-V_2(t')]$,
 and  with $\Delta= \frac{\Gamma_0}{\pi a}$, the  mean backscattering current is found from Eq.~(\ref{bsc}) as
\be \label{bscurr}
J_{\rm bs}(t)= \nu e \int_0^{\infty}d\tau~k_{\rm P}(\tau)\sin\big[G(t,t-\tau)\big]  \; .
\ee
Hence, time average of  the powers $P_{1/2}(t)$ given in Eq.~(\ref{Pexcess})  
with the backscattering current (\ref{bscurr}) directly yields the expression (\ref{S12}) for $i=1,\,2$, 
which coincides with the expression (1) of the Letter. In accordance with this, the power fluctuation resulting from backscattering are found as given in Eq.~(\ref{S14})  with (\ref{kDsm}), and in Eq.~(2) of the Letter.
Thus we have demonstrated complete correspondence in the scaling limit of the above QPC with the QBP system.
The virtue of the QPC geometry is, that powers running through the left- and right 
terminals resulting from the backscattering current, and the associated power fluctuations, 
can be measured individually.


\begin{thebibliography}{77}
\bibitem{remark}
The power law form (\ref{eq:kappaas}) in the range $\beta\omega \gg 1$ originates from the short-time behavior
$k_{\rm P}(\tau)\propto (\tau/\beta)^{-2K}$ of the expression (\ref{eq:ks}) holding in the Ohmic scaling limit.
 Hence, Ohmic bath coupling is ultimately responsible for the possible undercut of the trade-off bound observed
in Figs.~\ref{fig:two}(b) and \ref{fig:three}(d).
\end{thebibliography}
\end{document}